\documentclass{aa}
\usepackage{graphics}
\usepackage{epsfig}
\usepackage{latexsym}
\def\deg       {$^{\circ}$}

\def \sig      {$\sigma$}
\def \gray     {$\gamma$-ray}
\def \grays    {$\gamma$-rays}

\def \flux     {10$^{-5}$cm$^{-2}$s$^{-1}$}
\def \fluxa    {10$^{-8}$cm$^{-2}$s$^{-1}$}

\begin{document}
\hyphenation{brems-strah-lung}

\thesaurus{   
               03                    
              (13.07.2;              
               11.01.2;              
               11.17.4 3C~273);      
              }

\title{A large high-energy gamma-ray flare from the blazar 3C~273}

\author{ W.~Collmar\inst{1},
         O.~Reimer\inst{1},
         K.~Bennett\inst{4},
         H.~Bloemen\inst{2},
         W.~Hermsen\inst{2},
         G.G.~Lichti\inst{1},
         J.~Ryan\inst{3}, 
         V.~Sch\"onfelder\inst{1},
         H.~Steinle\inst{1}, 
         O.R.~Williams\inst{4}
         and M.~B\"ottcher\inst{5}
}
 
\institute{Max-Planck-Institut f\"ur extraterrestrische Physik,
               P.O. Box 1603, D-85740 Garching, Germany 
            \and
             SRON-Utrecht, Sorbonnelaan 2, NL-3584 CA Utrecht,
               The Netherlands
            \and
               University of New Hampshire, Institute for the Study
               of Earth, Oceans and Space, Durham NH 03824, USA
            \and
              Astrophysics Division, Space Science Department of
               ESA/ESTEC, NL-2200 AG Noordwijk, The Netherlands
            \and
              Rice University, Space Physics and Astronomy Department,
              6100 S. Main Street, Houston, TX 77005 -- 1892, USA
          }

\offprints{W.~Collmar}
\mail{wec@mpe.mpg.de}

\date{Received 30 July 1999 / Accepted 23 November 1999}

\titlerunning{A large high-energy gamma-ray flare from the blazar 3C~273}
\authorrunning{W.~Collmar et al.}
\maketitle

\begin{abstract}
The Compton Gamma-Ray Observatory (CGRO) experiments EGRET and COMPTEL
observed the Virgo sky region continuously for 7 weeks between December 10, 1996 and January 28, 1997. The prominent quasar 3C~273 was found to be the brightest source in \grays\ and was significantly detected by EGRET and COMPTEL. The EGRET experiment observed a time-variable flux at energies above 100~MeV, which reached in a 2-week flaring period (December 30, 1996 to January 14, 1997) its highest flux level observed during the CGRO-era. COMPTEL, however, does not observe obvious time variability at energies below $\sim$30~MeV contemporaneous to EGRET. In particular, no flare was observed, indicating that this outburst is solely a high-energy ($>$100~MeV) phenomenon. The energy spectrum between 3~MeV and 10~GeV is well represented by a simple power-law model. Below 3~MeV a spectral turnover is indicated. Performing spectral analysis for different time periods, we found evidence for a spectral hardening during the flaring period, which is consistent with the flare occurring mainly at the higher energies and with its absence at COMPTEL energies of a few MeV. 
This may be interpreted as an indication that the emission
in the EGRET energy range is dominated by a different radiation mechanism
than the MeV emission. We argue that the most likely mechanism for the
high-energy flare is inverse-Compton scattering of reprocessed accretion-disk
radiation.
\end{abstract}

\section{Introduction}
Since its discovery as an extragalactic object in 1962 (\cite{Schmidt63}),
the quasar 3C~273 is one of the best studied Active Galactic Nuclei (AGN).
With a redshift of 0.158 ($\sim$800 Mpc for H$_0$~=~60 km/s/Mpc)
it is relatively close, and by being bright in all wavelength
regions from radio to \gray\ energies, it is an excellent candidate for multiwavelength studies.  
 
After six years of operation, the EGRET experiment aboard CGRO has now detected \gray\ emission from more than 70 AGN at energies above 100~MeV (e.g. \cite{Hartman99}). These observations have dramatically changed our picture of these sources.    
With the exception of Cen~A (\cite{Sreekumar99}), 
all of them are identified with 
blazars (e.g. \cite{Mattox97}), the AGN subgroup consisting of either
flat-spectrum radio quasars or BL~Lacertae objects. Two remarkable \gray\ characteristics of these sources are 
that 1) they are highly variable down to time scales of a day or even 
shorter, and 2) that during flaring states the \gray\ luminosity can dominate
their bolometric power. 

The quasar 3C~273 is one of these blazar-type \gray\ loud AGN.  
It was first detected at \grays\ by the COS-B satellite at energies above 50~MeV (\cite{Swanenburg78}), and -- until the launch of CGRO in 1991 -- remained the only identified extragalactic point source at these energies.
3C~273 was redetected at \gray\ energies by the EGRET experiment in 1991 (\cite{Montigny93}). Analysing the first four years of EGRET data, \cite{Montigny97} found a time-variable \gray\ flux, consisting of detections as well as non-detections in individual observational periods. Spectral variability was observed as well, showing the trend of spectral hardening with increasing flux. The third EGRET source catalogue (\cite{Hartman99})
lists 3C~273 with an average flux value of (15.4$\pm$1.8) $\times$ 10$^{-8}$~ph~cm$^{-2}$~s$^{-1}$ for energies above 100~MeV and
the time period between April '91 and October '95.  

3C~273 was first discovered to be an emitter of low-energy \grays\ by COMPTEL in 1991 (\cite{Hermsen93}). The source is frequently detected in individual CGRO pointings (e.g. \cite{Collmar99}), however, non-detections occur as well proving time variability of the MeV-flux on time scales of months (\cite{Williams95}). In time-averaged analyses 3C~273 is detected very significantly and shows in
the 0.75-30~MeV band a soft spectrum, i.e. photon index $\alpha >$~2   (E$^{-\alpha}$) in combined data (\cite{Collmar96}). However, combining contemporaneous high-energy data reveals that the MeV-band is a transition region for the spectrum of 3C~273 showing a turnover from a harder ($\alpha \sim$1.7) spectrum at hard X-ray energies to a softer one ($\alpha \sim$2.5) at high-energy ($>$100~MeV) \grays\ (e.g. \cite{Lichti95}, von Montigny et al. 1997).     

At hard ($>$50~keV) X-rays 3C~273 is always significantly detected by the Oriented Scintillation Spectrometer Experiment (OSSE) showing flux 
variations in the 50-150~keV band up to a factor of 8 during 5~years (\cite{NaronBrown97}). 
A power-law spectrum with a photon index of roughly 1.7 (\cite{Johnson95}) up to $\sim$1~MeV is typically observed.
Above $\sim$1~MeV a spectral softening is found, consistent with the results of the multiwavelength campaigns. Recently, during the highest flux state as observed by OSSE, evidence for a low-energy spectral break at about 0.3~MeV was found (\cite{NaronBrown97}) suggesting an anticorrelation between flux
and break energy.    

In this paper we report on 7 weeks of continuous \gray\ observations by the EGRET and COMPTEL experiments aboard CGRO in December 1996 and January 1997. 
In Sect. 2 we describe the observations and the data analyses, in Sect. 3 we give the results, and discuss their implications in Sect. 4. Finally, 
the  conclusions are presented in Sect. 5.

\section{Observations and data analysis}
During Cycle~6 (October 15, 1996 to November 11, 1997) of its mission, CGRO was pointed continuously to the Virgo sky region for 7 weeks, beginning on December 10, 1996 and ending on January 28, 1997. The main target was the blazar 3C~279 which is located at ($\alpha , \delta)_{2000}$ = (12$^h$56$^m$11$^s$,  -5.8\deg47'21.5''), at a distance of
$\sim$10.5\deg\ from 3C~273. The relevant observational parameters are given in Table~1. 

\begin{table}[tbh]
\caption{Continuous CGRO observations of 3C~273 during Cycle~6. The viewing period (VP) number in CGRO-notation, the observational periods in calendar date and Truncated Julian Day (TJD), and the angular separation (Sep.) between 3C~273 and the pointing direction are given. The analysed time periods (VP-combinations) and their durations are given as well. 
For their explanation see text.
}
\begin{flushleft}
\begin{tabular}{cccc}
\hline\noalign{\smallskip}
VP  & \multicolumn{2}{c}{Observation Time}        & Sep. \\
    & yy/mm/dd - yy/mm/dd &  TJD           & [\deg] \\
\noalign{\smallskip}
\hline
 606.0 & 96/12/10 - 96/12/17 & 10427 - 10434 & 11.2 \\
 607.0 & 96/12/17 - 96/12/23 & 10434 - 10440 & 11.2 \\
 608.0 & 96/12/23 - 96/12/30 & 10440 - 10447 & 11.2 \\
 609.0 & 96/12/30 - 97/01/07 & 10447 - 10455 & 11.2 \\
 610.0 & 97/01/07 - 97/01/14 & 10455 - 10462 & 11.2 \\
 610.5 & 97/01/14 - 97/01/21 & 10462 - 10469 &  9.5  \\
 611.1 & 97/01/21 - 97/01/28 & 10469 - 10476 & 11.1 \\
\noalign{\smallskip}
\hline
       &                     &   &      \\
Periods &  VPs & Obs. Time   & Dur.     \\
        &      &  TJD        & [days]   \\
\noalign{\smallskip}
\hline
 All   & 606-611.1   & 10427 - 10476 & 49  \\ 
  A    & 606-608     & 10427 - 10447 & 20  \\ 
  B    & 609+610     & 10447 - 10462 & 15  \\ 
  C    & 610.5+611.1 & 10462 - 10476 & 14  \\ 
  D    &  A + C      &               & 34  \\
\noalign{\smallskip}
\hline
\end{tabular}\end{flushleft}
\label{tab1}
\end{table}

The spark-chamber telescope EGRET covers the energy range from
$\sim$30~MeV to $\sim$30~GeV. The instrument and its
calibration is described in detail
by \cite{Thompson93} and \cite{Esposito99}. 
The analysis of the EGRET data followed
the standard EGRET procedure i.e. using count and exposure maps as well as predictions of the diffuse \gray\ background (e.g. \cite{Hunter97}). 
The maps containing events with energies above 100~MeV were used for source detection and determination of the source position, and the ones 
of the 10 standard energy intervals for determination of the source 
spectrum by assuming a power law. 
The analysis applied the standard maximum-likelihood method (\cite{Mattox96})
and spatial selections ($<$25\deg\ off axis).   
Empirical flux correction were applied for energies below 70~MeV.  

\begin{figure*}[thb]
   \begin{picture}(180,60)(0,0)   
\put(0,0){\makebox(85,0)[lb]{\psfig{file=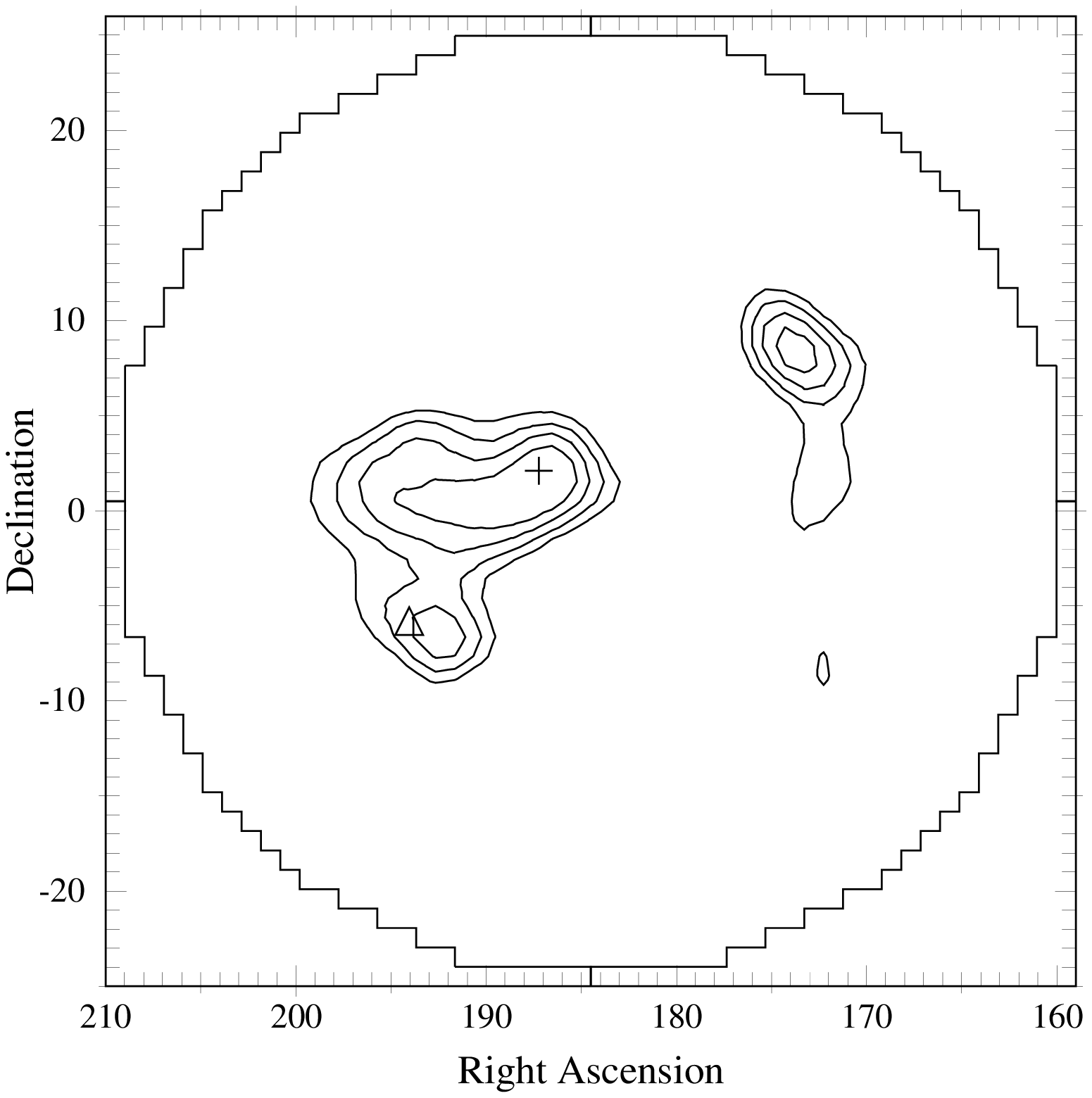,height=6.0cm,clip=}}}
\put(60,0){\makebox(85,0)[lb]{\psfig{file=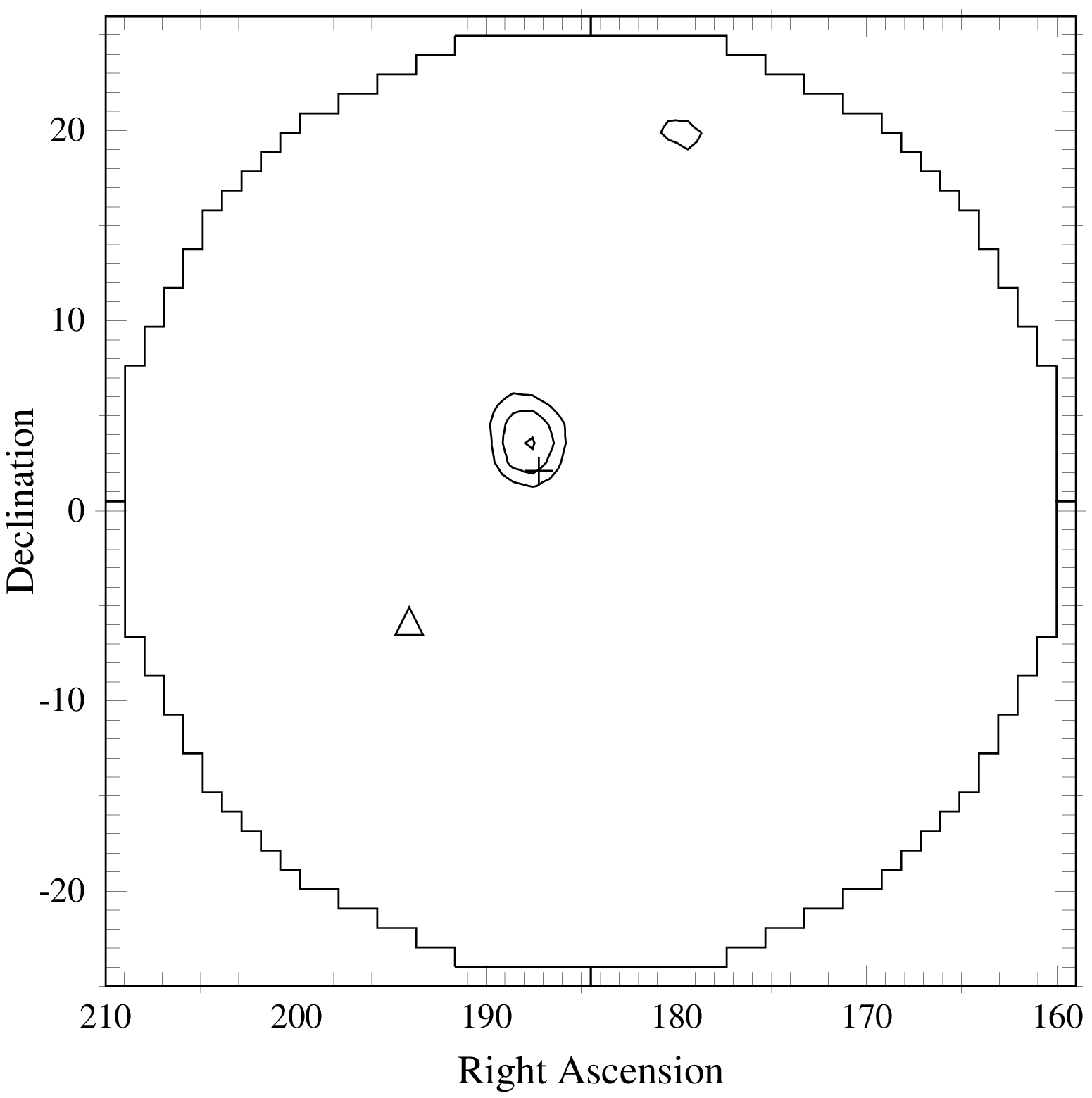,height=6.0cm,clip=}}}
\put(120,0){\makebox(85,0)[lb]{\psfig{file=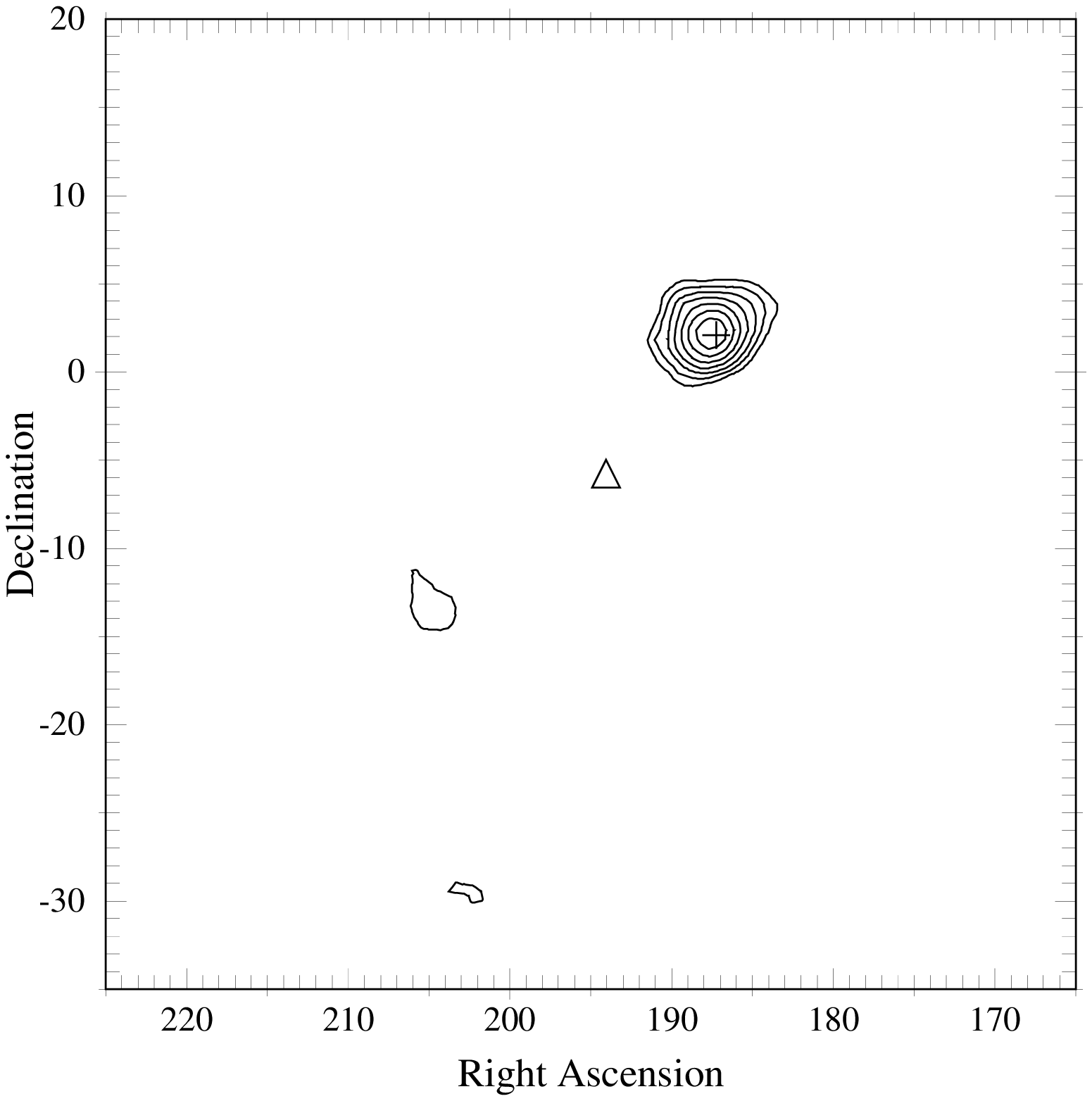,height=6.0cm,clip=}}}
    \end{picture}
\caption{Virgo maps in the \gray\ energy bands 3-10~MeV (left), 10-30~MeV (middle) as observed by COMPTEL, and $>$100~MeV (right) 
as provided by EGRET for the complete set of 7 weeks of continuous data (VPs 606 to 611.1). For COMPTEL the contour lines start at a detection significance of 3$\sigma$ ($\chi^2_1$-statistics for a known source) with a step of
0.5$\sigma$. 
For EGRET the contour lines also start at a detection significance of
3$\sigma$, however with a step of 1$\sigma$. 
The locations of the famous Virgo blazars 3C~273 (+) and 3C~279 ($\Delta$) are
indicated. 3C~273 is detected with 10.4\sig\ by EGRET. There is significant emission from 3C~273 in both COMPTEL maps. The COMPTEL 3-10~MeV map shows hints for additional sources. }
\label{fig1}
\end{figure*} 
%

The imaging Compton telescope COMPTEL covers the energy band $\sim$0.75 to $\sim$30~MeV.
For a detailed description of the COMPTEL instrument see \cite{Schonfelder93}.
 The COMPTEL data have been analysed following the COMPTEL standard maximum-likelihood analysis procedures, which for point-source analyses including background generation are described in sufficient detail by \cite{Bloemen94} and which derive quantitative source parameters like 
detection significances, fluxes, and flux errors. 
For consistency checks maximum-entropy images (see e.g. \cite{Strong92})
have been generated as well. 

The fluxes of 3C~273 were determined by simultaneously fitting further sources or source candidates which are indicated by the maps (see Fig.~\ref{fig1}).
This approach leads iteratively to simultaneous flux determinations of several
sources, including the generation of a background model which takes into
account the possible presence of further sources. 
This analysis has been carried out for the four standard COMPTEL energy bands. 
The flux results given in Sect.~3 have been derived with point spread functions
 assuming an E$^{-2.0}$ power-law shape for the sources, which is approximately the correct shape for the MeV-spectrum of 3C~273. 

\section{Results}

\subsection{Detections}
The EGRET data analysis of the combined 7 weeks of Virgo data revealed 
a strong source consistent with the location of 3C~273 (Fig.~1). The overall
detection significance at the position of the quasar for energies above
100~MeV is 10.4$\sigma$ assuming $\chi^2_1$-statistics for a known source.
The 7-week mean flux (E$>$100~MeV) is (43.4$\pm$5.8) $\times$ 10$^{-8}$~ph~cm$^{-2}$~sec$^{-1}$,  
which is roughly 3 times the average flux listed in the third EGRET catalogue (\cite{Hartman99}).
 A flux level in excess of that  --  (48.3$\pm$11.8) $\times$ 10$^{-8}$~ph~cm$^{-2}$~sec$^{-1}$ in VP~308.6 -- had been reported previously
only once (\cite{Hartman99}).
3C~273 is identified with the \gray\ source on the basis of its sky location. 

Simultaneously with these EGRET findings,
COMPTEL observed significant emission from
the sky position
of 3C~273 in three (1-3~MeV, 3-10~MeV, 10-30~MeV) out of its 4 standard
energy bands. The source is not detected at the lowest COMPTEL energies (0.75-1~MeV). The overall detection 
significance is 7.7$\sigma$. The average flux values in the 
individual COMPTEL bands (see Table~2) are among the largest 
ever measured in this energy window, showing that the source was active 
in \grays\ during this period.    

\begin{figure}[tbh]
\resizebox{\hsize}{!}{\includegraphics{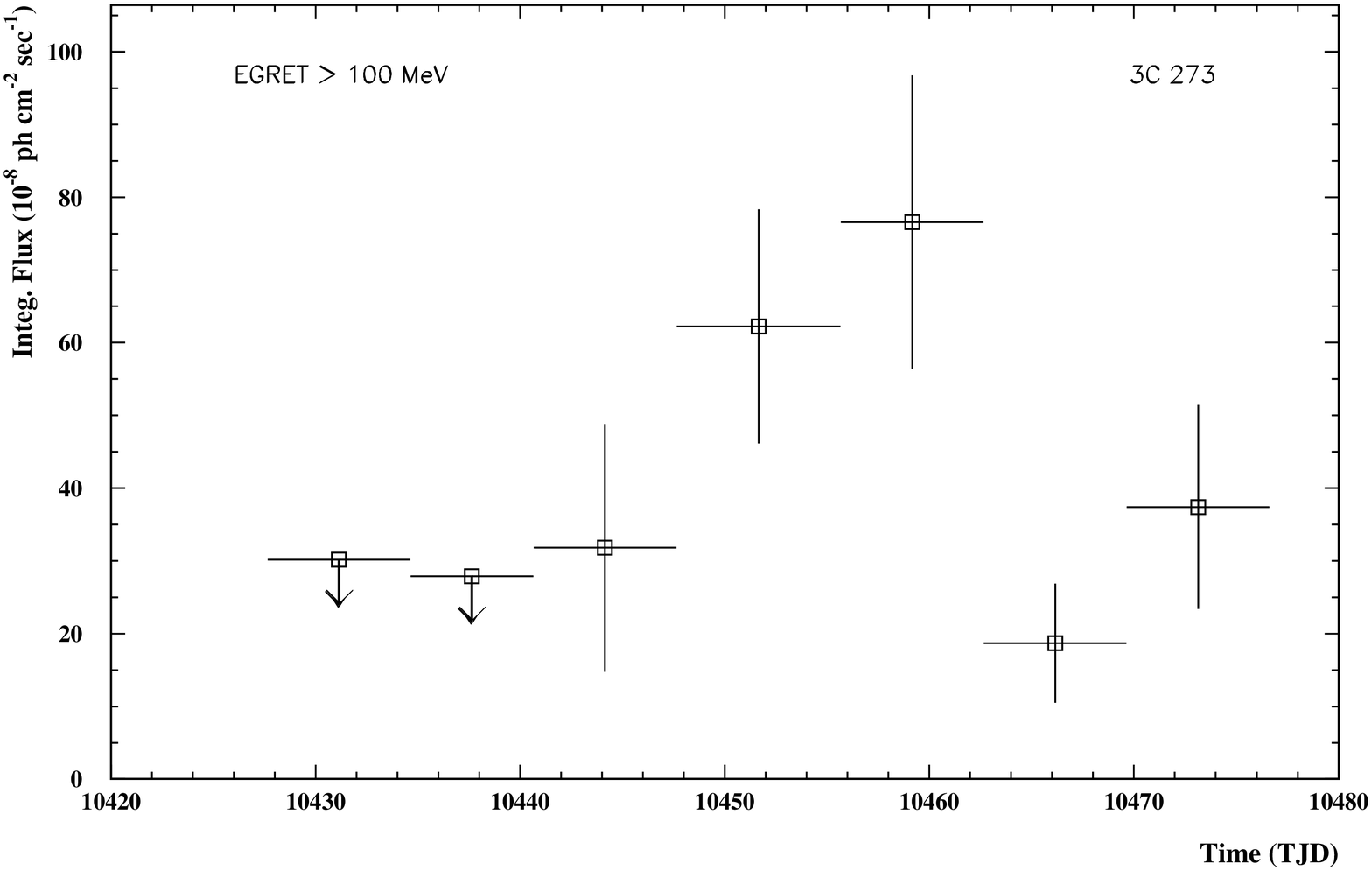}}
\caption{The EGRET light curve of 3C~273 at energy above 100~MeV for the seven individual VPs (for the calendar dates see Table~1). The quasar was not detected during the beginning of the observations, then showed a flux increase to the highest level ever for $\sim$2 weeks in the middle part (VPs 609 and 610), and then returned to about the previous flux level in the last two weeks.  
The errors are 1$\sigma$ and the upper limits are 2$\sigma$.}
\label{fig2}
\end{figure} 

\begin{figure}[tbh]
\resizebox{\hsize}{!}{\includegraphics{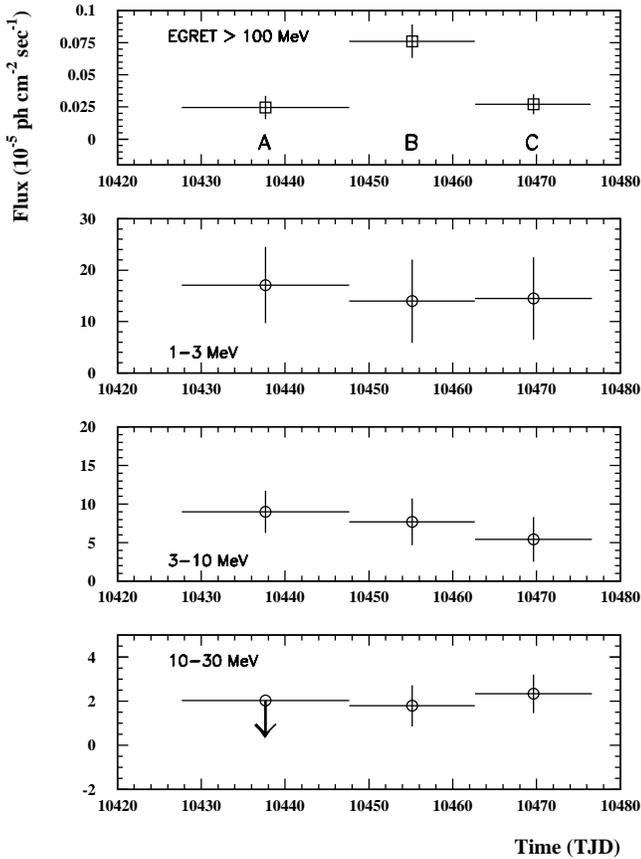}}
\caption{EGRET ($\Box$) and COMPTEL ($\circ$) light curves of 3C~273 in different \gray\ bands.
The definition of the 3 time periods (A, B, C) is given in the text.  
While a flare is clearly seen above 100~MeV by EGRET,
no obvious flux increase is observed at energies below 30~MeV by COMPTEL.
The errors are 1$\sigma$ and the upper limits are 2$\sigma$.
} 
\label{fig3}
\end{figure} 

\subsection{Time variability}
To check for time variability we have subdivided the total 7 week observation  into slices of the 7 individual VPs covering typically one week each (Table~1).
3C~273 was not detected by EGRET during the first
two VPs (detection-significance threshold 3.5\sig$\!$).
It then appeared just above the detection threshold and increased further
to show the largest \gray\ flare observed during the CGRO era. 
This maximum flux level of (77$\pm$20) $\times$ 10$^{-8}$~ph~cm$^{-2}$~sec$^{-1}$
is reached in VP~610. 
Thereafter the flux returned to an intermediate level, which is clearly 
detectable. Fits of the two EGRET light curves shown in Figs.~\ref{fig2}
and \ref{fig3} (upper panel)
assuming a constant flux resulted in $\chi^2_{min}$-values of 19.9 for the
light curve containing 
7 flux points (individual VPs) and 12.5 for the one containing 3 flux points. 
According to $\chi^2$-statistics these values correspond to probabilities of 
2.9~$\times$~10$^{-3}$ and 1.9~$\times$~10$^{-3}$, respectively, for a constant flux, which convert to 3.0\sig\ and 3.1\sig\ evidence for a time-variable 
$>$100~MeV flux.  
The largest change in flux occurred between VPs 610 and 610.5
when the flux dropped 
by a factor of $\sim$4.1 within 7~days. This is the shortest time variability 
at \gray\ energies ever reported for 3C~273.
The significance that the two flux values are different is 2.7$\sigma$. During the two-week outburst (VPs 609 and 610) the source reached a flux level even slightly in excess to the COS-B flux reported by Swanenburg et al. (1978).    

\begin{table}[bht]
\caption{Fluxes and upper limits for 3C~273 for the different analysed
periods in units of \flux\ for the COMPTEL bands and \fluxa\ for the 
EGRET band. The energy bands are given in MeV. The errors are 1$\sigma$. The upper limits are 2$\sigma$. An upper limit is given when the significance of an individual flux value is less than 1$\sigma$. The errors and upper limits are statistical only. The observational periods A-D are defined in Table~1 and described in the text.} 
\begin{flushleft}
\begin{tabular}{cccccc}
\hline\noalign{\smallskip}
   & \multicolumn{4}{c}{COMPTEL} & \multicolumn{1}{c}{EGRET} \\
Period & 0.75-1 & 1-3 & 3-10 & 10-30  & $\geq$100 \\
\hline\noalign{\smallskip}
All & $<$10.7 & 15.2$\pm$4.5 & 7.5$\pm$1.7 & 1.5$\pm$0.5 & 43.4$\pm$5.8 \\
 A  & $<$17.3 & 17.1$\pm$7.4 & 9.0$\pm$2.7 & $<$2.0      & 24.6$\pm$9.0 \\
 B & $<$23.0 & 14.0$\pm$8.1 & 7.7$\pm$3.0 & 1.8$\pm$0.9 & 76.1$\pm$12.9 \\
 C & $<$18.8 & 14.5$\pm$8.0 & 5.4$\pm$2.9 & 2.3$\pm$0.9 & 27.1$\pm$7.7 \\
 D & $<$12.8 & 15.8$\pm$5.4 & 7.4$\pm$2.0 & 1.4$\pm$0.6 & 27.7$\pm$6.0 \\
\hline\noalign{\smallskip}
\end{tabular}\end{flushleft}
\label{tab2}
\end{table}

We checked for time variability in the COMPTEL bands by subdividing the 
data into the same time slices as chosen for EGRET. No obvious time variability 
is visible in either energy band, however, the statistics in the
different COMPTEL bands became marginal, resulting in large error bars on the 
flux values. To improve the statistical significance we combined individual 
VPs. We defined three time intervals 
 which were selected according to 
the EGRET light curve: a pre-flare period (VPs 606-608) covering 3 weeks which we shall call A in the following, a flare period (VPs 609 and 610) covering two weeks which we shall call B, and a post-flare period 
(VPs 610.5 and 611) which we shall call C (see also Table~1). 
The simultaneous EGRET and COMPTEL 3C~273 fluxes for different energy bands
and various time periods are listed in Table~2 and are plotted in Fig.~\ref{fig3}. In contrast to EGRET, COMPTEL observes no obvious time variability. The same $\chi^2$-procedure as applied to the EGRET data 
showed that the different COMPTEL flux values are consistent with a constant
level as is also obvious from Fig.~\ref{fig3}. 
In particular the COMPTEL light curves do not show a hint of increased \gray\ emission during the EGRET flaring period. This result
suggests that the observed flare is either solely a high-energy ($>$30~MeV) 
phenomenon, or a time offset of at least 2 weeks between the EGRET and COMPTEL \gray\ bands is required. 

\subsection{Energy spectra}
The EGRET spectral analysis followed the standard EGRET procedure (see Sect.~2).
The energy range between 30~MeV and 10~GeV was subdivided into 10 energy intervals and the likelihood analysis was used to estimate the number of source photons in each energy bin. The data were fit to a single power-law model
of the following form 

\begin{equation}
I(E) = I_{0}  (E/E_{0})^{-\alpha} \,\, {\rm photons\,\,cm}^{-2} {\rm s}^{-1} {\rm MeV} ^{-1}
\end{equation}  

with the parameters $\alpha$ (photon spectral index) and $I_{0}$ (intensity at the normalization energy $E_{0}$). $E_{0}$ was chosen such, that the two free parameters are minimally correlated. 
We derived 1$\sigma$-errors on the parameters by adding 1.0 to the minimum $\chi^{2}$-value (\cite{Mattox96}). 

This approach was applied to the sum of all data as well as to selected  
subsets (see Fig.~\ref{fig3}, Table~1) to check for a
possible trend in time. In addition we summed the subsets A and C having 
roughly equal flux levels, which we shall call D, to check for a possible spectral trend with flux by using the improved event statistics.   
The results of the spectral fitting are given in Table~3.
First of all the spectra are well fitted by simple power-law functions: the average spectral index in the EGRET band is $\alpha$ = 2.40$\pm$0.14, 
which is comparable to previous results. For instance, \cite{Montigny97} found
spectral indices in the range between 2.2 and 3.2 with a trend of 
spectral hardening with increased source flux. The EGRET fits show this trend as well: the spectrum is hardest during the flaring period. However, this is not significant because the error on the spectral slope is quite large.   

\begin{figure}[tb]
\resizebox{\hsize}{!}{\includegraphics{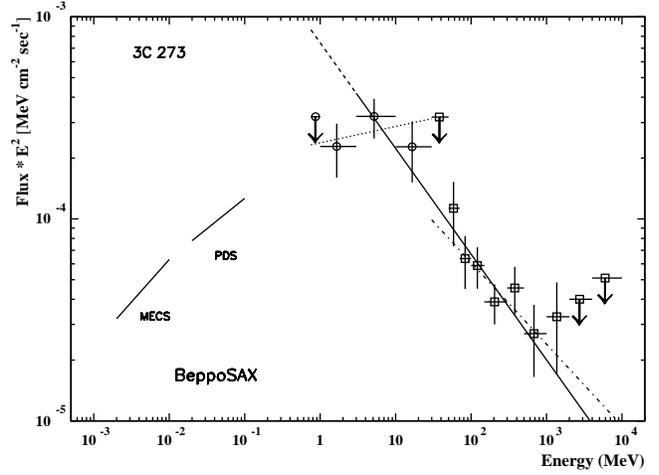}}
\caption{Quasi-simultaneous BeppoSAX-CGRO high-energy spectrum 
of 3C~273 in an E$^2 \times$ differential-flux representation. The EGRET
($\Box$) and COMPTEL ($\circ$) spectral points are derived from the sum of the whole observation (7~weeks). The errors are 1$\sigma$ and the upper limits are 2$\sigma$. An upper limit is drawn when the significance of an individual flux value is less than 1$\sigma$. The solid line represents the best-fit power-law model for the range 3~MeV to 10~GeV. The dashed lines show the extrapolation towards lower energies. The dotted line represents the best-fit power-law model
for only the COMPTEL data (0.75-30~MeV) and the dashed-dotted line the one for 
solely the EGRET data (30~MeV - 10 GeV). The BeppoSAX power-law shapes were observed on January 13, 1999 (\cite{Haardt98}), located within the CGRO observational period. The strong spectral turnover at a few MeV is evident. }
\label{fig4} 
\end{figure}

To derive the COMPTEL fluxes of 3C~273, we have applied the standard
maximum-likelihood method as described in Sect.~2. Background-subtracted and deconvolved source fluxes in the 4 standard energy bands have been derived by taking 
into account the presence of further \gray\ sources (3C~279 at $\alpha$/$\delta$ = 194.1\deg/-5.8\deg, 4C~-02.55 (1229-021) at  $\alpha$/$\delta$ = 188.0\deg/-2.4\deg)
and source candidates (at $\alpha$/$\delta$ = 193.5\deg/0.5\deg and  $\alpha$/$\delta$ = 173.5\deg/8.5\deg),
showing some evidence in the maps (Fig.~\ref{fig1}).
We note that inclusion of these 
further sources and source candidates has only a marginal effect on the flux
of 3C~273. Their inclusion or exclusion changes the derived 3C~273 fluxes only within its error bars.  
Assuming a power-law shape, the quasar is fitted typically with a spectral 
index of $\sim$2 throughout the COMPTEL energy band. For example, the sum of all data ('ALL') which has the best statistics, yields $\alpha$ = 1.92$\pm$0.13 between 0.75 and 30~MeV, which is significantly harder than $\alpha$ = 2.41$\pm$0.14 as found in the EGRET range for the same observational period (Fig.~\ref{fig4}). This result indicates a spectral 
hardening towards lower energies with the turnover starting at a few MeV. 
This spectral behaviour of 3C~273 is well known and has been reported previously
(e.g. \cite{Lichti95}, von Montigny et al. 1997).

\begin{figure}[tb]
\resizebox{\hsize}{!}{\includegraphics{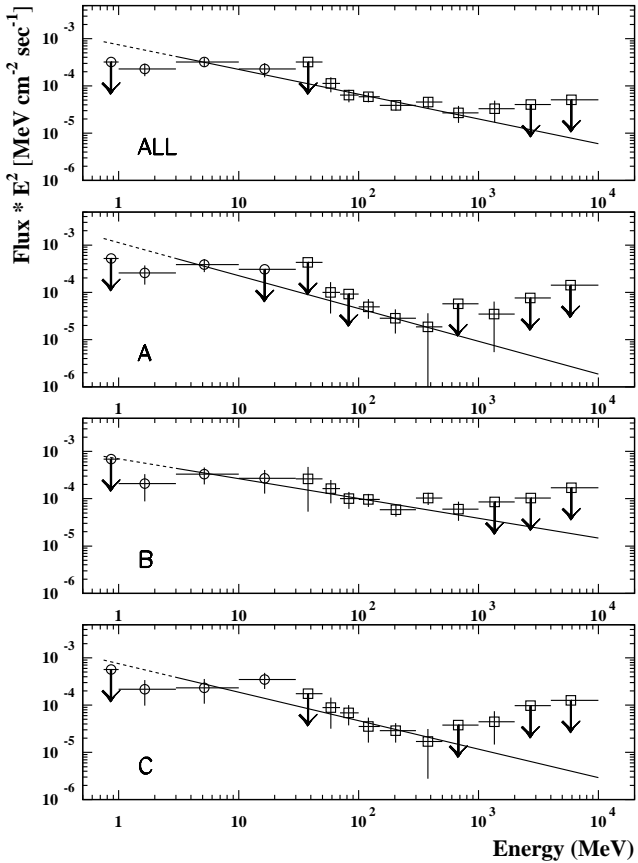}}
\caption{Combined EGRET ($\Box$) and COMPTEL ($\circ$) energy spectra for different time periods in an E$^2 \times$ differential-flux representation. The errors are 1$\sigma$ and the upper limits are 2$\sigma$. An upper limit is drawn when the significance of an individual flux value is less than 1$\sigma$. The solid line represents the best-fit power-law model for the range 3~MeV to 10~GeV. The dashed lines show the extrapolation towards lower energies.
}
\label{fig5}
\end{figure} 

To take advantage of the contemporaneous observations of both instruments
 in neighbouring energy regimes, we combined 
the deconvolved EGRET and COMPTEL spectra for the different time periods. 
Fitting the whole energy range (0.75~MeV to 10 GeV) with a power-law model, we derive harder spectral slopes and increased $\chi^2_{red}$-values compared 
to fitting solely the EGRET range. This effect weakens if we exclude the lowest-energy (0.75-1~MeV) spectral point, and disappears when we subsequently remove the next (1-3~MeV) COMPTEL flux point from the fitting procedure.
This behaviour is easily explained by the spectral turnover at low energies, which seems to affect the fits only below 3~MeV. 
Fitting a broken power-law model
\begin{equation}
I(E) = \left\{ \begin{array}{ll}
 I_{0}  (E/E_{0})^{-\alpha_{2}} &\mbox {if $E>E_{b}$} \\
 I_{0}  (E_{b}/E_{0})^{-\alpha_{2}} \, (E/E_{b})^{(\Delta\alpha - \alpha_{2})}
                                 &\mbox {if $E<E_{b}$}
\end{array}
\right.
\end{equation}  
where I$_0$ describes the differential source flux at the normalization
energy E$_{0}$, $\alpha_2$ the high-energy spectral photon index, 
$\Delta\alpha$ the break in spectral photon index towards lower energies
($\Delta\alpha$ = $\alpha_2$ - $\alpha_1$), and E$_{b}$ the break energy,  
provides consistent results. The best-fit value for 
the break energy for the sum of all data is found to be at $\sim$5~MeV, 
which, however, is not well defined due to the small lever-arm towards 
lower energies. Considering these facts, we conclude, that the EGRET 
power-law spectrum extends into the COMPTEL band 
down to $\sim$3~MeV before it is substantially altered by the spectral
turnover. This is illustrated in Fig.~\ref{fig4}. 
By chance the quasar was observed simultaneously in the X-ray band by the 
BeppoSax satellite. Haardt et al. (1998) published the X-ray results,
i.e. fluxes and spectra, of 3C~273 covering a monitoring period of 4 days (January 13, 15, 17, and 22, 1997). These X-ray observations 
are coincident in time with the transition from the \gray\ flaring period
to the moderate post-flare \gray\ level (Fig.~\ref{fig2}).
Fig.~\ref{fig4} shows a broad-band high-energy spectrum of 3C~273, containing
the COMPTEL and EGRET spectra for the sum of all 7 weeks and the best-fit 
power-law shape for the quasi-simultaneous X-ray measurements (covering only 
one day) provided by Haardt et al. (1998).   

\begin{table*}[thb]
\caption[]{Results of the power-law fitting (I(E) = I$_{0}$  (E/E$_{0})^{-\alpha}$) for the different time periods. The left part gives the results for fitting only the EGRET data (30~MeV - 10~GeV), while the right part shows the results of fitting the combined EGRET and COMPTEL (3~MeV - 10~GeV).  The errors on the fit parameters are derived by the $\chi^{2}_{min}$+1 contour
level. }
\begin{flushleft}
\begin{tabular}{ccccccccc}
\hline\noalign{\smallskip}
 & \multicolumn{4}{c}{EGRET (30~MeV-10~GeV)} & \multicolumn{4}{c}{COMPTEL + EGRET (3~MeV-10~GeV)} \\ 
Obs. & PL-Index & I$_0 \times 10^{-9}$ & E$_0$ & $\chi^{2}_{red}$ & 
PL-Index & I$_0 \times 10^{-9}$ & E$_0$ & $\chi^{2}_{red}$ \\
Period & ($\alpha$) & [ph/(cm$^{2}$ s MeV)] & [MeV] &  & 
($\alpha$) & [ph/(cm$^{2}$ s MeV)] & [MeV] &  \\ 
\hline\noalign{\smallskip}
All & 2.41$\pm$0.14  & 1.18$\pm$0.13 & 197.93 & 0.42 
    & 2.52$\pm$0.07  & 6.69$\pm$0.67 & 100    & 0.45 \\  
A   & 2.65$\pm$0.47  & 1.57$\pm$0.46 & 150.35 & 0.36 
    & 2.70$\pm$0.14  & 4.56$\pm$1.27 & 100    & 0.36 \\  
B   & 2.41$\pm$0.21  & 2.56$\pm$0.38 & 176.16 & 0.62 
    & 2.42$\pm$0.10 & 10.10$\pm$1.36 & 100    & 0.52 \\  
C   & 2.52$\pm$0.32 & 1.04$\pm$0.25 & 175.65  & 0.41 
    & 2.60$\pm$0.13  & 4.70$\pm$1.04 & 100    & 0.61 \\  
D   & 2.61$\pm$0.27  & 1.30$\pm$0.24 & 160.65 & 0.60 
    & 2.66$\pm$0.10  & 4.67$\pm$0.75 & 100    & 0.52 \\  
\hline\noalign{\smallskip}
\end{tabular}\end{flushleft}
\label{tab3}
\end{table*}

\begin{figure}[tbh]
\resizebox{\hsize}{!}{\includegraphics{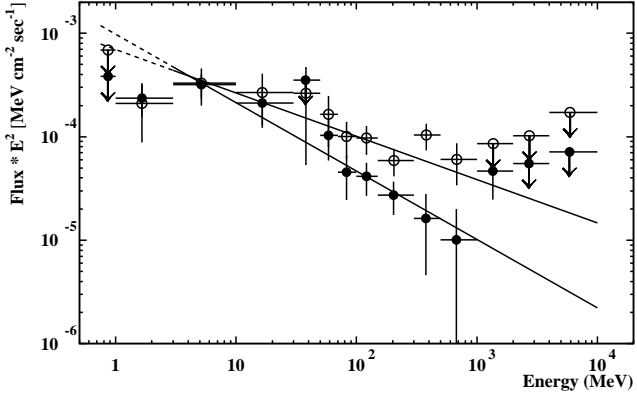}}
\caption{The spectra of the flare state (B, open circles) and the sum of the lower flux levels (A+C, filled circles) are compared. The two spectra differ mainly above 100~MeV.}
\label{fig6}
\end{figure}

\begin{figure}[tbh]
\resizebox{\hsize}{!}{\includegraphics{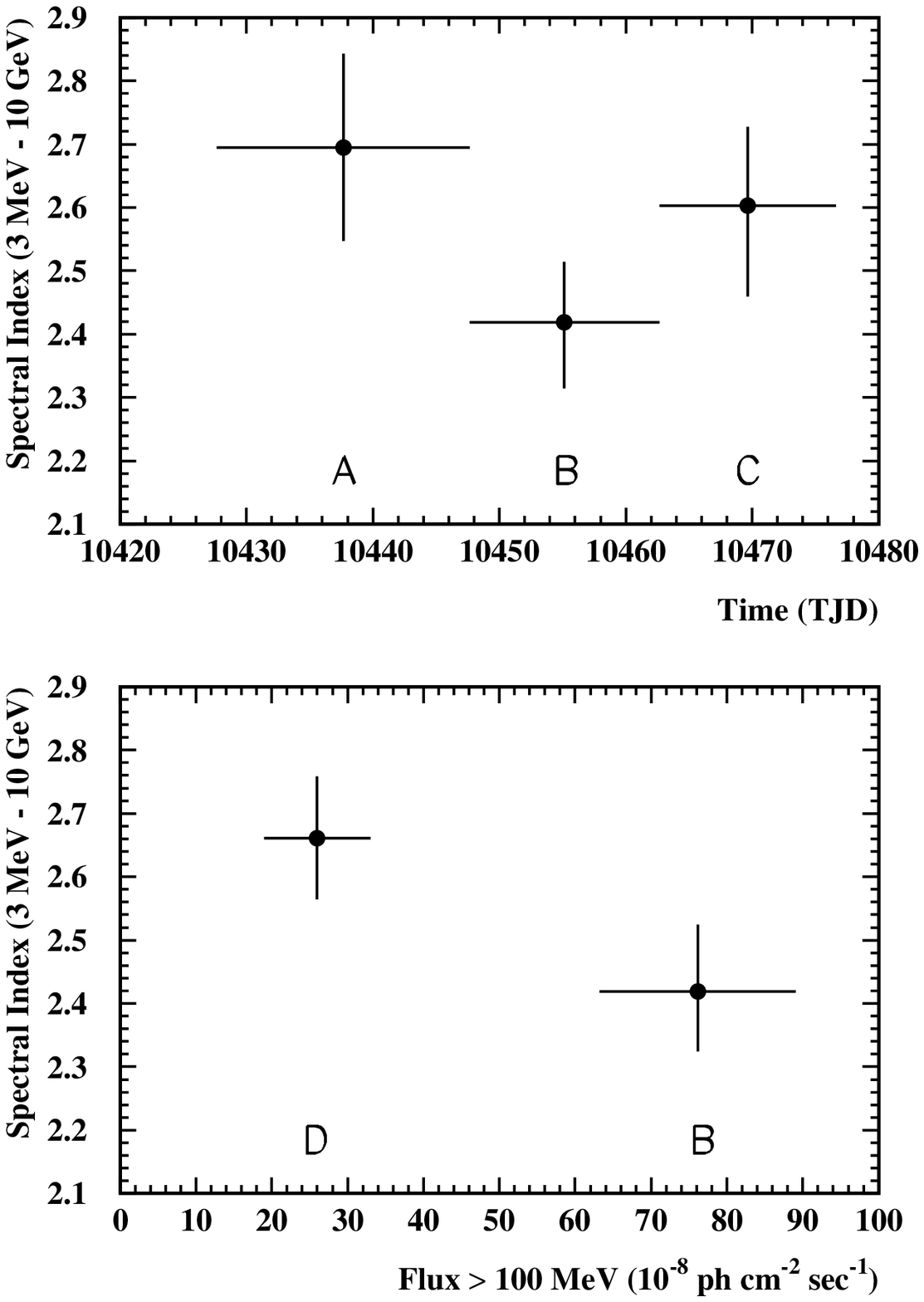}}
\caption{ The spectral index (3~MeV - 10~GeV) as a function of time  (periods A, B, and C) is shown in the upper panel and of source flux $>$100~MeV
(periods B and D (A+C)) in the lower panel. The errors are 1$\sigma$. There is evidence for spectral hardening with increasing flux.}
\label{fig7}
\end{figure}

To investigate this high-energy component in more detail, we fitted 
the different observational subsets between 3~MeV and 10~GeV with single 
power-law models (Fig.~\ref{fig5}).
We take advantage of this enlarged (with respect to only EGRET) energy band,
for which the spectral index can be determined 
more accurately.
 So, for the 3~MeV to 10~GeV band, the trend that during the flare the spectrum hardens as suggested by the EGRET analysis (see above), 
is observed more significantly. Especially, if we compare the periods B and D.
The 1$\sigma$ statistical errors in spectral index during the flare and non-flare intervals do not overlap anymore. 
The fit results for the EGRET band only and this enlarged band are given in Table~3, and the latter ones are shown graphically in the Figs.~\ref{fig5}
and \ref{fig6}. 
Along the 7-week observation, 3C~273 is observed to have a steep spectrum 
at the beginning, which hardens during the two-week flaring period, and 
turns back to roughly the same shape in the 2 week post flare period 
(Fig.~\ref{fig7}). This result is consistent with the constant flux  
observed at COMPTEL energies. The flare occurs mainly at energies 
above 100~MeV, not affecting the COMPTEL points which results in a hardening 
of the overall \gray\ spectrum. 

From this spectral analysis, we conclude too, 
that we either observed a phenomenon which occurs solely at high \gray\  
energies, or
that there are time delays between the different \gray\ bands. The first case would require an additional spectral component triggered by some mechanism which is only effective at EGRET energies. The second possibility would require that the COMPTEL energies are either delayed by two weeks or would be in advance
by three weeks with respect to EGRET. 

\section{Discussion}
At the end of 1996 and early 1997 the CGRO experiments EGRET and COMPTEL
observed the well-known quasar 3C~273 at \gray\ energies continuously for 7 weeks.
The blazar was \gray\ active and therefore it was significantly detected
by both experiments. 
Assuming isotropic emission, H$_0$=60~km/s/Mpc, and a q$_0$=0.5 cosmology,
we derive an average 7-week luminosity in the EGRET band (100~MeV - 10~GeV) of $\sim$1.7$\times$10$^{46}$erg/s, for the flaring period (period B in Table~2) 
$\sim$2.7$\times$10$^{46}$erg/s, and $\sim$0.9$\times$10$^{46}$erg/s for the
periods outside the two-week flare. The 7-week average 
luminosity in the COMPTEL band (1-30~MeV) is derived to be  $\sim$12$\times$10$^{46}$erg/s. 

The blazar was simultaneously observed in the X-ray band
by the BeppoSax satellite in a monitoring fashion covering a period of
1.5~weeks with four individual pointings.
In X-rays the quasar was about 15\% brighter in the first observation than
in the last one (\cite{Haardt98}), which is consistent with the 
\gray\ behaviour above 100~MeV. In addition, they note that during these monitoring observations the source was, on average, a factor of 2 brighter as observed half a year earlier. This suggests a correlation of X- and 
\gray\ behaviour of 3C~273. 

Simultaneous X-ray and \gray\ measurements provide the possibility of
estimating some physical source parameters, if one assumes that both photon 
populations are generated co-spatially. This is not an unreasonable assumption, given the indication of correlated variability as mentioned above. 
It was observed also in other blazars like 3C~279 for example (\cite{Wehrle98}).
Using the simultaneous X-ray spectra of the BeppoSax Medium Energy Concentrator
Spectrometers (MECS) published by Haardt et al. (1998) we derive a flux of $\sim$16$\mu$Jy at 1~keV for 3C~273. Applying the expression for the lower
limit on the Doppler factor, $\delta$, given by 
\cite{Do_Gh95} and assuming H$_0$=60~km/s/Mpc, we derive 

\begin{equation}
\delta \geq \left( 684 \, t^{-1}_{var} \, E_{\gamma}^{\alpha} 
\right)^{\frac{2}{(4 + 2 \alpha)}}, 
\end{equation}

where t$_{var}$ is the variability time scale in days, E$_{\gamma}$ the highest unabsorbed \gray\ energy in GeV, and $\alpha$ the spectral index in X-rays.
With a variability scale of one week as observed by EGRET
 at a significance level of 2.7\sig, 
an E$_{\gamma}$ of 1~GeV, and an $\alpha$ of 0.6 (energy index) as observed 
simultaneously by the BeppoSax MECS,
we derive a lower limit on the Doppler factor of 

\begin{equation}
\delta \geq 2.4  .
\label{dopp}
\end{equation}

A $\delta \geq$ 2.4 implies that the \gray\ luminosities mentioned above overestimates the intrinsic luminosities at least by a factor
of about 33 since $L_{obs} = \delta^{3 + \alpha} L_{intr}$. 

Although EGRET observed time variability at energies above 100~MeV,
the COMPTEL experiment between 1 and 30~MeV simultaneously
measures a constant flux of 3C~273. In particular,  
COMPTEL observes no hints of increased \gray\ emission in any of its energy
bands during the two-week flaring period. This is a surprising result, and 
in some respects different to simultaneous COMPTEL/EGRET observations
of other flaring blazars. For example, during the major outburst of 3C~279, the 
COMPTEL 10-30~MeV flux followed the flux trend as observed by EGRET at 
higher energies (\cite{Collmar97a}). Also, by analysing the COMPTEL 
data of the first 3.5 years on the blazar PKS~0528+134, \cite{Collmar97b} 
found the trend that the COMPTEL upper energy band follows the EGRET 
light curve, while the emission in the COMPTEL 1-3~MeV band was independent 
of the EGRET-observed behaviour. According to the measurements presented here,
the largest \gray\ flux ($>$100~MeV) of 3C~273 is only
a high-energy phenomenon, 
because it is (at least simultaneously) restricted  to energies above 
$\sim$30~MeV. This is consistent with the hardening of the \gray\ spectrum 
during the flare. This observation suggests either an additional spectral
component which becomes important at energies above 100~MeV 
or a time-offset between the high- and low-energy \grays\ is required. 

A different behaviour in the MeV- and $>$100~MeV band had been observed 
in the so-called 'MeV-blazars' GRO~J0506-609 (\cite{Bloemen95}) and 
PKS~0208-512 (\cite{Blom95}), which - in contrast to the presented case -
showed simultaneously strong MeV-emission compared to weak emission 
above 100~MeV. These sources indicated first that several emission 
components or mechanisms can be operating at \gray\ energies.    

In the standard models, the \gray\ emission is generated within a jet, 
where blobs, filled with relativistic leptons, are moving at relativistic
speeds along the jet axis. The \gray\ emission is generated by inverse-Compton interactions of these blob leptons with soft photons which either 
are provided by the environment (e.g. accretion disk) of the jet or are 
self-generated synchrotron photons. 
In such a picture, the observed behaviour of 3C~273 could be qualitatively explained
by a change in the energy distribution of the blob leptons, by a change of 
the energy distribution of the soft target photons or by both. 
If the \gray\ flare is triggered by a change in the energy distribution of the 
blob leptons, the observation would require that only the high-energy end of 
the distribution, responsible for the \grays\ in the EGRET band, is increased
in energy as well as in number density, while at lower energies the
distribution has to remain constant to keep the MeV-emission unchanged. 
This case, if applicable, might provide hints on the particle acceleration 
mechanism. If a variation of the soft photon distribution is 
responsible for the \gray\ flare, a flare in a certain wavelength band, 
e.g. UV flare of accretion disk photons, could trigger the event. 

An apparently natural explanation of this uncorrelated EGRET-COMPTEL 
behaviour would be if the MeV- and $>$100~MeV-emissions emanate from spatially 
different regions. However, we consider this explanation unlikely because 
-- as mentioned above -- simultaneous \gray\ variations have been 
observed by EGRET and COMPTEL in several blazars. Also during the observational
period presented here, both experiments found 3C~273 in an active
\gray\ state, which suggests
a common region of photon generation.  

In our opinition the most plausible scenario appears to be that the MeV 
and the GeV emissions are dominated by different radiation mechanisms with 
different flaring amplitudes, but emitted co-spatially by the same population
of relativistic electrons which are also responsible at least for the 
high-frequency part of the synchrotron component. Such a two-component
scenario for spectral variability during high-energy flares has first 
been suggested for PKS~0528+134 by \cite{Collmar97b}; see also
\cite{Bo_Co98} and \cite{Mu99}. Given typical parameter values 
for the relativistic electron distribution in AGN jets, and scaling
$\delta$ in units of 3 (due to the results of Eq.~\ref{dopp}),
the synchrotron emission peaks at

\begin{equation}
\langle\epsilon\rangle_{Sy} \sim 7 \cdot 10^{-8} B_0
\langle\gamma\rangle_3^2 \left( {\delta \over 3} \right),
\label{e_sy}
\end{equation}
where $\epsilon = h \nu / (m_e c^2)$ is the dimensionless photon
energy, $B_0$ is the co-moving magnetic field in Gauss, and $\langle
\gamma\rangle_3$ is the average Lorentz factor of the electrons
in units of $10^3$. With $B_0$ and $\langle\gamma\rangle_3$ being 
of order unity, the synchrotron peak is at $\nu_{sy} \sim 10^{13}$~Hz, 
as generally observed for 3C~273. We will discuss two possible radiation 
mechanisms for the observed high-energy flare: a) Comptonization of 
accretion disk radiation which is reprocessed by broad-line region clouds 
(ECC for External Comptonization of radiation from Clouds;   
\cite{Sikora94}), and b) Comptonization of synchrotron radiation from the jet, 
reprocessed by broad-line region clouds (RSy for synchrotron reflection) 
as proposed by \cite{Ghi_Ma96}. The ECC spectrum, which assumes 
the accretion disk ('blue bump') photons to be the soft target photons,
is expected to have a rather narrow spectral distribution, peaking around

\begin{equation}
\langle\epsilon\rangle_{ECC} \sim 3 \cdot 10^3
\langle\epsilon_D\rangle_{-4} \Gamma_{10}
\langle\gamma\rangle_3^2 \left( {\delta \over 3} \right),
\label{e_ecc}
\end{equation}
where $\langle\epsilon_D\rangle_{-4}$ ($\equiv \langle\epsilon_D\rangle$/10$^{-4}$)  -- being of order unity -- is the dimensionless $\nu F_{\nu}$ peak energy of the accretion disk spectrum and $\Gamma_{10}$ is the bulk Lorentz factor in units of 10, which is also assumed
to be of order unity.
Consequently, this peak is expected to be at $\sim 1$~GeV. 
In contrast, the RSy spectrum is expected to be much broader (similar to the
synchrotron self-Compton spectrum) and peaks around

\begin{equation}
\langle\epsilon\rangle_{RSy} \sim 7 B_0
\langle\gamma\rangle_3^4 \Gamma_{10}^{2}
\left( {\delta \over 3} \right),
\label{e_rsy}
\end{equation}
typically at MeV energies. The factors $\Gamma_{10}$ and $\Gamma_{10}^{2}$
in Eqs. (\ref{e_ecc}) and (\ref{e_rsy}) arise from the Lorentz boost
of the external radiation field into the comoving rest frame of the
blob and of the jet synchrotron radiation into the stationary frame
of the BLR.

The estimates derived from Eqs. (\ref{e_ecc}) and (\ref{e_rsy}) indicate
that a flare in the EGRET energy range without a significant
variation of the MeV emission is more likely to be caused by
the ECC mechanism than by the RSy scenario.
We point out that these interpretations are based on time-variability of EGRET which is signifcant at 3.1\sig.  
 
Recently \cite{McHardy99} reported simultaneous mm, infrared (IR), and X-ray (3-20~keV) observations of 3C~273 which cover in part the same time
period as the \gray\ observations reported here.
In particular they observed a simultaneous IR and X-ray flare lasting for 
about 10~days, which in fact is simultaneous to the high-energy \gray\ 
flare observed by EGRET.
These simultaneous flares add important information in light of the two-component hypothesis for the \gray\ spectrum proposed here and 
provide additional constraints for the modelling of the emission
processes in this quasar.  
A discussion and interpretation of the peculiar variability pattern of
3C 273 in IR, X-rays and \grays\ as observed in early 1997 will be given 
by \cite{Bo_Co00}.  

\section{Conclusion}
From December 10, 1996 to January 28, 1997 the CGRO instruments EGRET
and COMPTEL observed the Virgo sky region continuously for 7 weeks, 
detecting 3C~273 in an active \gray\ state.
EGRET ($>$100~MeV) observed a time-variable flux, peaking during a 2-week flaring period at its highest level observed during the CGRO-era.
COMPTEL, however, does not observe any contemporaneous \gray\ flare at
energies below $\sim$30~MeV, 
showing that this outburst is restricted to \gray\ energies above 30~MeV.
This is consistent with the spectral hardening observed
in the 3~MeV to 10~GeV energy band during the flaring period. 
 
The peculiar variability properties of the flare may be explained
in terms of a two-component spectral model with
the emission in the EGRET energy range produced by Comptonization
of reprocessed accretion disk emission.
The different variability behaviour in \grays\ is inconsistent with the synchrotron-reflection model being the cause of the \gray\ flare.

This observation covers an opportune sequence of low pre-flare,
high flare, and again low post-flare emission in \grays.
In general, this \gray\ observation could turn out
to be important for further modelling of blazar emission processes 
because the \gray\ flare is well located in time and therefore can
possibly be correlated to flux measurements of monitoring observations
in other wavelength regions.

\acknowledgements
This research was supported by the German government through DLR grant 50 QV 9096 8, by NASA under contract NAS5-26645, and by the Netherlands
Organisation for Scientific Research NWO.


\begin{thebibliography}{}

\bibitem[Bloemen et al.\ (1994)]{Bloemen94}
Bloemen H., Hermsen W, Swanenburg B.N., et al., 1994, ApJS 92, 419

\bibitem[Bloemen et al.\ 1995]{Bloemen95}
Bloemen H., Bennett K., Blom J.J., et al., 1995, A\&A 293, L1

\bibitem[Blom et al.\ 1995]{Blom95}
Blom J.J., Bloemen H., Bennett K., et al., 1995, A\&A 295, 330

\bibitem[B\"ottcher \& Collmar\ (1998)]{Bo_Co98}
B\"ottcher M., Collmar W., 1998, A\&A 329, 57

\bibitem[B\"ottcher \& Collmar\ (2000)]{Bo_Co00}
B\"ottcher M., Collmar W., 2000, ApJ, submitted

\bibitem[Collmar et al.\ 1996]{Collmar96}
Collmar W., Bennett K., Bloemen H., et al., 1996, A\&AS 120, 515

\bibitem[Collmar et al.\ 1997a]{Collmar97a}
Collmar W., Bennett K., Bloemen H., et al., 1997a, in Proceedings of the 
Fourth Compton Symposium, eds. C.D. Dermer, M.S. Strickman, J.D. Kurfess (New York: AIP Conf. Proc. 410), p. 1341  

\bibitem[Collmar et al.\ (1997b)]{Collmar97b}
Collmar W., Bennett K., Bloemen H., et al., 1997b, A\&A 328, 33

\bibitem[Collmar et al.\ 1999]{Collmar99}
Collmar W., Bennett K., Bloemen H., et al., 1999, Proc. of 3rd INTEGRAL
Workshop 'The Extreme Universe', eds. G. Palumbo, A. Bazzano, C. Winkler, 
Astrophys. Letters and Communications, in press 

\bibitem[ Dondi \& Ghisellini\ (1995)]{Do_Gh95}
Dondi L., Ghisellini G. 1995, MNRAS 273, 583

\bibitem[Esposito et al.\ (1999)] {Esposito99}
Esposito J.A., Bertsch D.L., Chen A.W., et al., 1999, ApJS 123, 203

\bibitem[Ghisellini \& Madau\ (1996)] {Ghi_Ma96}
Ghisellini G., Madau P., 1996, MNRAS 280, 67

\bibitem[Haardt et al.\ 1998] {Haardt98}
Haardt F., Fossati G., Grandi P., et al., 1998, A\&A 340, 35

\bibitem[Hartman et al.\ 1999] {Hartman99}
Hartman R.C., Bertsch D.L., Bloom S.D., et al., 1999, ApJS 123, 79

\bibitem[Hermsen et al.\ 1993] {Hermsen93}
Hermsen W., Aarts H.J.M., Bennett K., et al., 1993, A\&AS 97, 97

\bibitem[Hunter et al.\ 1997] {Hunter97}
Hunter S.D., Bertsch D.L., Catelli J.R., et al., 1997, ApJ 481, 205

\bibitem[Johnson et al.\ 1995] {Johnson95}
Johnson W.N., Dermer C.D., Kinzer R.L.,  et al., 1995, ApJ 445, 182

\bibitem[Lichti et al.\ 1995]{Lichti95}
Lichti G. G., Balonek T., Courvoisier T.J.-L., et al., 1995,  A\&A 298, 711

\bibitem[Mattox et al.\ 1996] {Mattox96}
Mattox J.R., Bertsch D.L., Chiang J., et al., 1996, ApJ 461, 396

\bibitem[Mattox et al.\ 1997] {Mattox97}
Mattox J.R., Schachter J., Molnar L., et al., 1997, ApJ 481, 95

\bibitem[McHardy et al.\ (1999)] {McHardy99}
McHardy I., Lawson A., Newsam A., et al., 1999, MNRAS, accepted

\bibitem[McNaron-Brown et al.\ 1997] {NaronBrown97}
McNaron-Brown K., Johnson W.N., Dermer C.D.,  et al., 1997, ApJ 474, L85

\bibitem[von~Montigny et al.\ 1993] {Montigny93}
 von~Montigny C., Bertsch D.L., Fichtel C.E., et al., 1993, A\&A Suppl. 97, 101

\bibitem[von~Montigny et al.\ (1997)] {Montigny97}
von~Montigny C., Aller H., Aller M., et al., 1997, ApJ 483, 161

\bibitem[Mukherjee et al.\ (1999)] {Mu99}
Mukherjee R., B\"ottcher M., Hartman R.C., et al., 1999, ApJ, in press

\bibitem[Schmidt 1963] {Schmidt63}
Schmidt M., 1963, Nature 197, 1040

\bibitem[Sch\"onfelder et al.\ (1993)]{Schonfelder93}
Sch\"{o}nfelder V., Aarts H., Bennett K., et al., 1993, ApJS 86, 657 

\bibitem[Sikora et al.\ 1994] {Sikora94}
Sikora M., Begelman M.C., Mitchell C., et al., 1994, ApJ 421, 153   

\bibitem[Sreekumar et al.\ 1999] {Sreekumar99}
Sreekumar P., Bertsch D.L., Hartman R.C., et al., 1999, Astropart. Physics 11, 221

\bibitem[Strong et al.\ 1992] {Strong92}
Strong A.W., Cabeza-Orcel P., Bennett K., et al., 1992, in Data Analysis in Astronomy IV, eds. V. Di Ges\`{u}, L. Scarsi, R. Buccheri, et al., (New York: plenum Press), p. 251

\bibitem[Swanenburg et al.\ 1978] {Swanenburg78}
Swanenburg B.N., Hermsen W., Bennett K., et al., 1978,  Nature 275, 298

\bibitem[Thompson et al.\ (1993)] {Thompson93}
Thompson D.J., Bertsch D.L., Fichtel C.E., et al., 1993, ApJS 86 , 629

\bibitem[Wehrle et al.\ 1998] {Wehrle98}
Wehrle A., Pian E., Urry C.M., et al., 1998, ApJ 497, 178

\bibitem[Williams et al.\ 1995]{Williams95}
Williams O.R., Bennett K., Bloemen H., et al., 1995, A\&A 298, 33


\end{thebibliography}
\end{document}